\documentclass[aps,prl,twocolumn,superscriptaddress,showpacs]{revtex4-1}
\RequirePackage{lineno}
\usepackage{graphicx} 
\usepackage{epstopdf}
\usepackage{amssymb}
\usepackage{lineno}
\usepackage{color}
\usepackage[normalem]{ulem}

\newcommand{\pom}{$\, \mathrm{I\!P} \, $}
\newcommand{\dpe}{D$\mathrm{I\!P}$E }

\newcommand{\pipi}{$\pi^+\pi^-$}

\def \nobreakseq {\nobreak \hskip 0pt \hbox}

\begin{document}


\lefthyphenmin=2
\righthyphenmin=3

\preprint{FERMILAB-PUB-15-033-PPD}

\title{Measurement of
central exclusive $\pi^+\pi^-$ production in $p\bar{p}$ collisions \\ at $\sqrt{s}$ = 0.9 and 1.96 TeV at CDF
}
\affiliation{Institute of Physics, Academia Sinica, Taipei, Taiwan 11529, Republic of China}
\affiliation{Argonne National Laboratory, Argonne, Illinois 60439, USA}
\affiliation{University of Athens, 157 71 Athens, Greece}
\affiliation{Institut de Fisica d'Altes Energies, ICREA, Universitat Autonoma de Barcelona, E-08193, Bellaterra (Barcelona), Spain}
\affiliation{Baylor University, Waco, Texas 76798, USA}
\affiliation{Istituto Nazionale di Fisica Nucleare Bologna, \ensuremath{^{mm}}University of Bologna, I-40127 Bologna, Italy}
\affiliation{University of California, Davis, Davis, California 95616, USA}
\affiliation{University of California, Los Angeles, Los Angeles, California 90024, USA}
\affiliation{Instituto de Fisica de Cantabria, CSIC-University of Cantabria, 39005 Santander, Spain}
\affiliation{Carnegie Mellon University, Pittsburgh, Pennsylvania 15213, USA}
\affiliation{Enrico Fermi Institute, University of Chicago, Chicago, Illinois 60637, USA}
\affiliation{Comenius University, 842 48 Bratislava, Slovakia; Institute of Experimental Physics, 040 01 Kosice, Slovakia}
\affiliation{Joint Institute for Nuclear Research, RU-141980 Dubna, Russia}
\affiliation{Duke University, Durham, North Carolina 27708, USA}
\affiliation{Fermi National Accelerator Laboratory, Batavia, Illinois 60510, USA}
\affiliation{University of Florida, Gainesville, Florida 32611, USA}
\affiliation{Laboratori Nazionali di Frascati, Istituto Nazionale di Fisica Nucleare, I-00044 Frascati, Italy}
\affiliation{University of Geneva, CH-1211 Geneva 4, Switzerland}
\affiliation{Glasgow University, Glasgow G12 8QQ, United Kingdom}
\affiliation{Harvard University, Cambridge, Massachusetts 02138, USA}
\affiliation{Division of High Energy Physics, Department of Physics, University of Helsinki, FIN-00014, Helsinki, Finland; Helsinki Institute of Physics, FIN-00014, Helsinki, Finland}
\affiliation{University of Illinois, Urbana, Illinois 61801, USA}
\affiliation{The Johns Hopkins University, Baltimore, Maryland 21218, USA}
\affiliation{Institut f\"{u}r Experimentelle Kernphysik, Karlsruhe Institute of Technology, D-76131 Karlsruhe, Germany}
\affiliation{Center for High Energy Physics: Kyungpook National University, Daegu 702-701, Korea; Seoul National University, Seoul 151-742, Korea; Sungkyunkwan University, Suwon 440-746, Korea; Korea Institute of Science and Technology Information, Daejeon 305-806, Korea; Chonnam National University, Gwangju 500-757, Korea; Chonbuk National University, Jeonju 561-756, Korea; Ewha Womans University, Seoul, 120-750, Korea}
\affiliation{Ernest Orlando Lawrence Berkeley National Laboratory, Berkeley, California 94720, USA}
\affiliation{University of Liverpool, Liverpool L69 7ZE, United Kingdom}
\affiliation{University College London, London WC1E 6BT, United Kingdom}
\affiliation{Centro de Investigaciones Energeticas Medioambientales y Tecnologicas, E-28040 Madrid, Spain}
\affiliation{Massachusetts Institute of Technology, Cambridge, Massachusetts 02139, USA}
\affiliation{University of Michigan, Ann Arbor, Michigan 48109, USA}
\affiliation{Michigan State University, East Lansing, Michigan 48824, USA}
\affiliation{Institution for Theoretical and Experimental Physics, ITEP, Moscow 117259, Russia}
\affiliation{University of New Mexico, Albuquerque, New Mexico 87131, USA}
\affiliation{The Ohio State University, Columbus, Ohio 43210, USA}
\affiliation{Okayama University, Okayama 700-8530, Japan}
\affiliation{Osaka City University, Osaka 558-8585, Japan}
\affiliation{University of Oxford, Oxford OX1 3RH, United Kingdom}
\affiliation{Istituto Nazionale di Fisica Nucleare, Sezione di Padova, \ensuremath{^{nn}}University of Padova, I-35131 Padova, Italy}
\affiliation{University of Pennsylvania, Philadelphia, Pennsylvania 19104, USA}
\affiliation{Istituto Nazionale di Fisica Nucleare Pisa, \ensuremath{^{oo}}University of Pisa, \ensuremath{^{pp}}University of Siena, \ensuremath{^{qq}}Scuola Normale Superiore, I-56127 Pisa, Italy, \ensuremath{^{rr}}INFN Pavia, I-27100 Pavia, Italy, \ensuremath{^{ss}}University of Pavia, I-27100 Pavia, Italy}
\affiliation{University of Pittsburgh, Pittsburgh, Pennsylvania 15260, USA}
\affiliation{Purdue University, West Lafayette, Indiana 47907, USA}
\affiliation{University of Rochester, Rochester, New York 14627, USA}
\affiliation{The Rockefeller University, New York, New York 10065, USA}
\affiliation{Istituto Nazionale di Fisica Nucleare, Sezione di Roma 1, \ensuremath{^{tt}}Sapienza Universit\`{a} di Roma, I-00185 Roma, Italy}
\affiliation{Mitchell Institute for Fundamental Physics and Astronomy, Texas A\&M University, College Station, Texas 77843, USA}
\affiliation{Istituto Nazionale di Fisica Nucleare Trieste, \ensuremath{^{uu}}Gruppo Collegato di Udine, \ensuremath{^{vv}}University of Udine, I-33100 Udine, Italy, \ensuremath{^{ww}}University of Trieste, I-34127 Trieste, Italy}
\affiliation{University of Tsukuba, Tsukuba, Ibaraki 305, Japan}
\affiliation{Tufts University, Medford, Massachusetts 02155, USA}
\affiliation{University of Virginia, Charlottesville, Virginia 22906, USA}
\affiliation{Waseda University, Tokyo 169, Japan}
\affiliation{Wayne State University, Detroit, Michigan 48201, USA}
\affiliation{University of Wisconsin, Madison, Wisconsin 53706, USA}
\affiliation{Yale University, New Haven, Connecticut 06520, USA}

\author{T.~Aaltonen}
\affiliation{Division of High Energy Physics, Department of Physics, University of Helsinki, FIN-00014, Helsinki, Finland; Helsinki Institute of Physics, FIN-00014, Helsinki, Finland}
\author{M.G.~Albrow}
\affiliation{Fermi National Accelerator Laboratory, Batavia, Illinois 60510, USA}
\author{S.~Amerio\ensuremath{^{nn}}}
\affiliation{Istituto Nazionale di Fisica Nucleare, Sezione di Padova, \ensuremath{^{nn}}University of Padova, I-35131 Padova, Italy}
\author{D.~Amidei}
\affiliation{University of Michigan, Ann Arbor, Michigan 48109, USA}
\author{A.~Anastassov\ensuremath{^{z}}}
\affiliation{Fermi National Accelerator Laboratory, Batavia, Illinois 60510, USA}
\author{A.~Annovi}
\affiliation{Laboratori Nazionali di Frascati, Istituto Nazionale di Fisica Nucleare, I-00044 Frascati, Italy}
\author{J.~Antos}
\affiliation{Comenius University, 842 48 Bratislava, Slovakia; Institute of Experimental Physics, 040 01 Kosice, Slovakia}
\author{G.~Apollinari}
\affiliation{Fermi National Accelerator Laboratory, Batavia, Illinois 60510, USA}
\author{J.A.~Appel}
\affiliation{Fermi National Accelerator Laboratory, Batavia, Illinois 60510, USA}
\author{T.~Arisawa}
\affiliation{Waseda University, Tokyo 169, Japan}
% ...

\collaboration{CDF Collaboration}
\altaffiliation[With visitors from]{
\ensuremath{^{a}}University of British Columbia, Vancouver, BC V6T 1Z1, Canada,
\ensuremath{^{b}}Istituto Nazionale di Fisica Nucleare, Sezione di Cagliari, 09042 Monserrato (Cagliari), Italy,
\ensuremath{^{c}}University of California Irvine, Irvine, CA 92697, USA,
\ensuremath{^{d}}Institute of Physics, Academy of Sciences of the Czech Republic, 182~21, Czech Republic,
\ensuremath{^{e}}CERN, CH-1211 Geneva, Switzerland,
\ensuremath{^{f}}University of Cologne, 50937 Cologne, Germany,
\ensuremath{^{g}}Cornell University, Ithaca, NY 14853, USA,
\ensuremath{^{h}}University of Cyprus, Nicosia CY-1678, Cyprus,
\ensuremath{^{i}}Deutsches Elektronen-Synchrotron DESY, 22607 Hamburg, Germany,
\ensuremath{^{j}}Office of Science, U.S. Department of Energy, Washington, DC 20585, USA,
\ensuremath{^{k}}University College Dublin, Dublin 4, Ireland,
\ensuremath{^{l}}ETH, 8092 Z\"{u}rich, Switzerland,
\ensuremath{^{m}}Institute for Nuclear Physics, Forschungszentrum J\"{u}lich GmbH, 52425 J\"{u}lich, Germany,
\ensuremath{^{n}}University of Fukui, Fukui City, Fukui Prefecture, Japan 910-0017,
\ensuremath{^{o}}Universidad Iberoamericana, Lomas de Santa Fe, M\'{e}xico, C.P. 01219, Distrito Federal,
\ensuremath{^{p}}University of Iowa, Iowa City, IA 52242, USA,
\ensuremath{^{q}}Kinki University, Higashi-Osaka City, Japan 577-8502,
\ensuremath{^{r}}Kansas State University, Manhattan, KS 66506, USA,
\ensuremath{^{s}}Brookhaven National Laboratory, Upton, NY 11973, USA,
\ensuremath{^{t}}Istituto Nazionale di Fisica Nucleare, Sezione di Lecce, Via Arnesano, I-73100 Lecce, Italy,
\ensuremath{^{u}}Queen Mary, University of London, London, E1 4NS, United Kingdom,
\ensuremath{^{v}}University of Melbourne, Victoria 3010, Australia,
\ensuremath{^{w}}Muons, Inc., Batavia, IL 60510, USA,
\ensuremath{^{x}}Nagasaki Institute of Applied Science, Nagasaki 851-0193, Japan,
\ensuremath{^{y}}National Research Nuclear University, Moscow 115409, Russia,
\ensuremath{^{z}}Northwestern University, Evanston, IL 60208, USA,
\ensuremath{^{aa}}University of Notre Dame, Notre Dame, IN 46556, USA,
\ensuremath{^{bb}}Universidad de Oviedo, E-33007 Oviedo, Spain,
\ensuremath{^{cc}}CNRS-IN2P3, Paris, F-75205 France,
\ensuremath{^{dd}}Universidad Tecnica Federico Santa Maria, 110v Valparaiso, Chile,
\ensuremath{^{ee}}The University of Jordan, Amman 11942, Jordan,
\ensuremath{^{ff}}Universite catholique de Louvain, 1348 Louvain-La-Neuve, Belgium,
\ensuremath{^{gg}}University of Z\"{u}rich, 8006 Z\"{u}rich, Switzerland,
\ensuremath{^{hh}}Massachusetts General Hospital, Boston, MA 02114 USA,
\ensuremath{^{ii}}Harvard Medical School, Boston, MA 02114 USA,
\ensuremath{^{jj}}Hampton University, Hampton, VA 23668, USA,
\ensuremath{^{kk}}Los Alamos National Laboratory, Los Alamos, NM 87544, USA,
\ensuremath{^{ll}}Universit\`{a} degli Studi di Napoli Federico I, I-80138 Napoli, Italy
}
\noaffiliation

\date{\today}

\begin{abstract}
We measure exclusive $\pi^+\pi^-$ production in proton-antiproton collisions at center-of-mass energies $\sqrt{s}$ = 0.9 and 1.96 TeV in the Collider Detector at Fermilab. We select events with two oppositely charged particles, assumed to be pions, with pseudorapidity $|\eta| < 1.3$ and with no other particles detected in $|\eta| < 5.9$. We require the \pipi system to have rapidity $|y|<$ 1.0. The production mechanism of these events is expected to be dominated by double pomeron exchange, which constrains the quantum numbers of the central state. The data are potentially valuable for isoscalar meson spectroscopy and for understanding the pomeron in a region of transition between nonperturbative and perturbative quantum chromodynamics. The data extend up to dipion mass $M(\pi^+\pi^-)$ = 5000 MeV/$c^2$ and show resonance structures attributed to $f_0$ and $f_2(1270)$ mesons. From the $\pi^+\pi^-$ and $K^+K^-$ spectra, we place upper limits on exclusive $\chi_{c0}(3415)$ production.
\end{abstract} 
\maketitle

In quantum chromodynamics, the theory of strong interactions between quarks and gluons, calculations of hadronic interactions are most reliable in the perturbative regime of high four-momentum transfer squared, i.e., for distance scales much less than the size of hadrons, typically 1 fm. Diffractive processes with low transverse-momentum ($p_T$)~\cite{kine} hadrons involve nonperturbative physics where Regge theory describes scattering processes~\cite{donnachie,acf}. 
The data presented in this paper, from proton-antiproton ($p\bar{p}$) collisions at $\sqrt{s}$ = 0.9 and 1.96 TeV, extend the experimental study of central exclusive production to above the charmonium threshold,  where the calculation of exclusive $\chi_c$ production by gluon fusion involves perturbative QCD processes \cite{superchic,szczchic}. Elastic scattering and other diffractive interactions are characterized by  a large region of rapidity \cite{kine}, $\Delta y$ (or $\Delta \eta$ as an approximation), devoid of hadrons, called a rapidity gap. Such processes are described in Regge theory by the exchange of a pomeron, \pom, which at leading order is a pair of gluons in a color-singlet state \cite{donnachie}. 

Central exclusive production is here defined to be $p  \bar{p} \rightarrow p^{(*)} \oplus X \oplus \; \bar{p}^{(*)}$, where $X$ is a specific central ($|y_X| < 1$) state and $\oplus$ represents a large region of rapidity, $1.3 < |\eta| < 5.9$, where no particles are detected. The incident particles remain intact or dissociate diffractively  ($p \rightarrow p^*$) into undetected hadrons. In this study we do not detect outgoing (anti)protons, and we  include events where they dissociate  into hadrons with $|\eta| > 5.9$. With two large rapidity gaps and central hadrons, the process is expected to be dominated by double pomeron exchange, \dpe \cite{donnachie,acf}.
Only at hadron colliders with center-of-mass energy $\sqrt{s} \gtrsim$ 50 GeV~\cite{afs, abcdhw} are rapidity gaps larger than $\Delta y$ = 3 possible with central state masses $M(X)$ up to about 2500 MeV/$c^2$. Calculations of the hadron mass spectrum in this domain have large uncertainties and do not yet include resonances. The CDF Collaboration reported the first observations of exclusive \pom + \pom $\rightarrow\gamma \gamma$ \cite{cdfgg}, and \pom + \pom $\rightarrow \chi_{c}$ using the  $J/\psi + \gamma$ decay mode \cite{cdfchic}, which can be calculated semiperturbatively through quark-loop diagrams \cite{durham,superchic,szczchic,hkms,cracow}.

In \dpe the central state $X$ must have isotopic spin $I = 0$ (isoscalar) with positive parity, C-parity and G-parity, and dominantly even spin J, so exclusive production of $f_0$,~$f_2$,~$\chi_{c0(2)}$, and $\chi_{b0(2)}$ mesons is allowed. Thus, \dpe is a \emph{quantum number filter}, favoring states having valence gluons, such as glueballs, i.e., hadrons with no valence quarks. Such states are expected in QCD, but 40 years after being proposed \cite{gellmann}, their existence is not established \cite{glueball}. 
More measurements in different production modes and decay channels should provide insight on the issue. In addition to its role in meson spectroscopy, \dpe studies shed light on the nature of the pomeron. Data at different collision energies provide additional tests of the theory; in Regge theory the cross section for $p + (\pi^+\pi^-) + \bar{p}$, with the $\pi^+\pi^-$ in a fixed central region, decreases approximately like 1/ln($s$) \cite{azimov,desai}.

The analysis reported here uses data from the CDF II detector, a general purpose detector to study $p\bar{p}$ collisions at the Fermilab Tevatron, and is described in detail elsewhere~\cite{cdf}. Here we give a brief summary of the detector components used in this analysis. Surrounding the beam pipe is a tracking system consisting of silicon microstrip detectors and a cylindrical drift chamber in a 1.4~Tesla solenoidal magnetic field.  The tracking system is close to 100\% efficient at reconstructing the trajectories of isolated charged particles with $p_T > 0.4$ GeV/$c$ and $|\eta|<1.3$. A barrel of time-of-flight (ToF) counters surrounds the drift chamber for $|\eta| \lesssim$ 0.9.
The magnet coil is surrounded by the central, end-wall ($|\eta| < 1.32$) and plug ($1.32 < |\eta|<3.64$) calorimeters.  These scintillator/photomultiplier sampling calorimeters have separate electromagnetic (EM) and hadronic (HAD) compartments with pointing tower geometry. Gas Cherenkov detectors (CLC) \cite{clc} covering 3.7 $<|\eta|<$ 4.7 monitor the luminosity and are used in veto to reject events with charged particles in that rapidity interval. Beam shower counters (BSC) consisting of 1.7 radiation lengths of lead followed by scintillation counters are located at $5.4<|\eta|<5.9$. The uninstrumented regions $4.7 < |\eta|< 5.4$ contribute to the nonexclusive background.
 
The data collected at $\sqrt{s}$ = 1.96 and 0.9 TeV correspond to integrated luminosities of 7.23 and 0.075 pb$^{-1}$, respectively, with a 6\% uncertainty. Only data-taking periods with $\lesssim 4$ inelastic collisions per bunch-crossing could be used. The 0.9 TeV data are from a special 40 hour period in 2011, with only three bunches of protons and antiprotons, and with low luminosity per bunch. The  first stage online event selection (level-1 trigger) requires two calorimeter towers (EM + HAD) in $|\eta| < 1.3$ to have $E_T \gtrsim$ 0.5 GeV, with a veto on any signals in the BSC and CLC counters. A higher-level trigger rejects events with any significant energy deposit in the plug calorimeter.  These requirements retain events with activity exclusively in the central region of the detector and reject most events with additional inelastic interactions in the same bunch crossing.  

We select events with two charged particles, each with $|\eta| < 1.3$ and $p_T >$ 0.4 ~GeV/$c$, and no other activity significantly above noise levels in the full detector, to $|\eta| =$ 5.9. The noise levels are determined for each subdetector using bunch-crossing (zero-bias) triggers in which no tracks or CLC hits are detected. We apply a requirement (cut) both on the sum of all signals in each subdetector and on the photomultiplier with the highest signal in each calorimeter. As we do not detect the final-state (anti)protons, the data include diffractive dissociation if all the produced hadrons have $|\eta| >$ 5.9, with higher $p^*$ masses allowed at $\sqrt{s}$ = 1.96 TeV than at 0.9 TeV.

As we only use bunch-crossings with no other visible interaction, we define an effective integrated luminosity $L_{\mathrm{eff}}$. This is determined from the probability that the full detector is empty, in zero-bias events, using the above noise criteria, as a function of the individual bunch luminosity $L_{\mathrm{bunch}}$. The distribution is exponential with the intercept consistent with 1.0 and slope consistent with the expected visible ($|\eta| <$ 5.9) fraction \cite{mbr} of the inelastic cross section \cite{totem}. We find $L_{\mathrm{eff}}$ = 1.16 (0.059) pb$^{-1}$ at $\sqrt{s}$ = 1.96 (0.9) TeV, with a 6.7\% uncertainty.

We assume the particles to be pions and discuss non-$\pi\pi$ backgrounds later. We require $|y(\pi\pi)| < 1.0$ and study differential cross sections $d\sigma/dM(\pi\pi)$ up to 5000 MeV/$c^2$. The charged particle tracks are required to have a good fit with $\geq$ 25 hits in both the stereo and axial layers of the drift chamber, with a $\chi^2$/DoF $< 2.5$, to both pass within 0.5 mm of the beam line in the transverse plane, and to be within 1 cm of each other in $z$ at their closest approach. This rejects cosmic ray background, nonprompt pairs (e.g. $K^0_S \rightarrow \pi^+\pi^-$) and events with poorly measured tracks. Each track is projected to the calorimeter, where it is required to deposit an energy that meets  the trigger requirements. We suppress events with neutral particles, or unreconstructed charged particles, by requiring no other energy deposits in the central calorimeters outside the cones $\sqrt{\Delta \eta^2 + \Delta \phi^2} =$ 0.3 around the 
extrapolated track positions.  

The events with same-sign hadrons are approximately 6.5\% of the total, and are rejected. These are nonexclusive events with at least two undetected charged particles, either due to an inefficiency or having very low $p_T$ with no reconstructed track and no calorimetric energy above the noise level.  

There is also a background from opposite-sign hadron pairs that are not $\pi^+\pi^-$. This is determined using the timing information from the ToF counters, available only when both particles have $|\eta| \lesssim$ 0.9. Only 67\% of all pairs have both particles identified as $\pi, K$, or $p$, and for these (89$\pm$1)\% are $\pi^+\pi^-$. As a check we restrict both tracks to have $|\eta| < 0.7$, and then 90\% of the pairs are identified; there is no significant change in the composition. All the spectra presented are for hadron pairs with assigned pion masses and include non-$\pi^+\pi^-$ backgrounds. The final sample contains 127 340 (6240) events at $\sqrt{s}$ = 1.96 (0.9) TeV with two opposite-charge particles in the chosen kinematic region, $p_T > 0.4$ GeV/$c$ and $|\eta| < 1.3$, and with $|y(\pi\pi)| < 1.0$. 

We present acceptance-corrected and normalized differential cross sections $d\sigma/dM(\pi\pi)$ in two kinematic regions: integrated over all $p_T(\pi\pi)$ for $M(\pi\pi) > 1000$ MeV/$c^2$ and integrated over $p_T(\pi\pi) > 1$ GeV/$c$ for $M(\pi\pi) > 300$ MeV/$c^2$. The region with smaller $p_T(\pi\pi)$  \emph{and} $M(\pi\pi)$ has limited acceptance and trigger efficiency. We calculate the acceptance and reconstruction efficiency by generating single pions, simulating the CDF detector response with a \textsc{geant-3} Monte Carlo program~\cite{geant}, and applying the selection criteria. This gives the four-dimensional product of geometrical acceptance, detection and reconstruction efficiencies, $A[p_T(\pi^+),p_T(\pi^-),\eta(\pi^+),\eta(\pi^-)]$, that we fit with an empirical smooth function. The trigger efficiency is obtained from
minimum-bias data, selecting isolated tracks and determining the probability that the towers hit by the particle fire the trigger.

To compute the event acceptance we generate states $X = \pi^+\pi^-$, uniform in rapidity over $|y(\pi\pi)| <$ 1.0, in [$M(\pi\pi),p_T(\pi\pi)$] bins, using a mass range $M(\pi\pi)$ from 2$m_\pi$ to 5000 MeV/$c^2$ with $p_T(\pi\pi)$ from 0 to 2.5 GeV/$c$ and with an isotropic $\pi^+\pi^-$ distribution in the $X$-frame. The data, binned in $M(\pi\pi)$ and $p_T(\pi\pi)$, are divided by the acceptance and $L_{\mathrm{eff}}$ to obtain the differential cross sections. The systematic uncertainty on the cross sections is dominated by the luminosity (6\%) and the choice of exclusivity cuts, which affect both the candidate event selection and $L_{\mathrm{eff}}$. These cuts are varied in the data over a reasonable range, and the resulting systematic uncertainty is shown in the plots as shading.

\begin{figure}
\centering
\includegraphics[width=86mm]{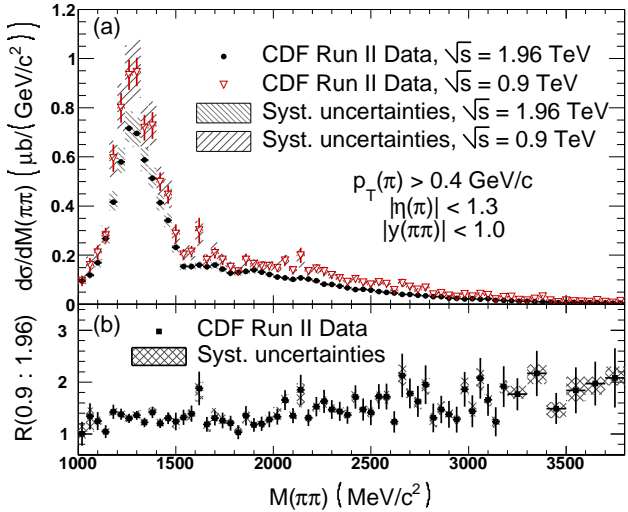}
\caption{(a) Differential cross section $d\sigma/dM(\pi\pi)$ for two charged particles, assumed to be $\pi^+\pi^-$, 
with $p_T > 0.4$ GeV/$c$, $|\eta| <$ 1.3 and $|y(\pi\pi)| < $ 1.0
 between two rapidity gaps $1.3 < |\eta| < 5.9$. Red open circles for $\sqrt{s}$ = 0.9 TeV and black points for $\sqrt{s}$ = 1.96 TeV. 
 (b) Ratio of cross sections at $\sqrt{s}$ = 0.9 and 1.96 TeV.}
\label{fig:Mass1960900}
\end{figure}

\begin{figure}
\centering
\includegraphics[width=86mm]{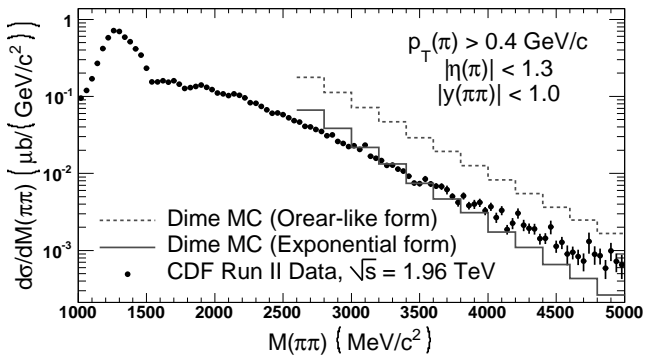}
\caption{Differential cross section $d\sigma/dM(\pi\pi)$ at $\sqrt{s}$ = 1.96 TeV for two charged particles, assumed to be $\pi^+\pi^-$, 
with $p_T > 0.4$ GeV/$c$, $|\eta| <$ 1.3 and $|y(\pi\pi)| < $ 1.0
 between two rapidity gaps $1.3 < |\eta| < 5.9$. Only statistical errors are shown; systematic uncertainties contribute approximately 10\% at all masses.
 The lines show predictions of Ref. \cite{vkpriv} with two different pion form factors.}
\label{fig:MassLog1960}
\end{figure}

\begin{figure}
\centering
\includegraphics[width=86mm]{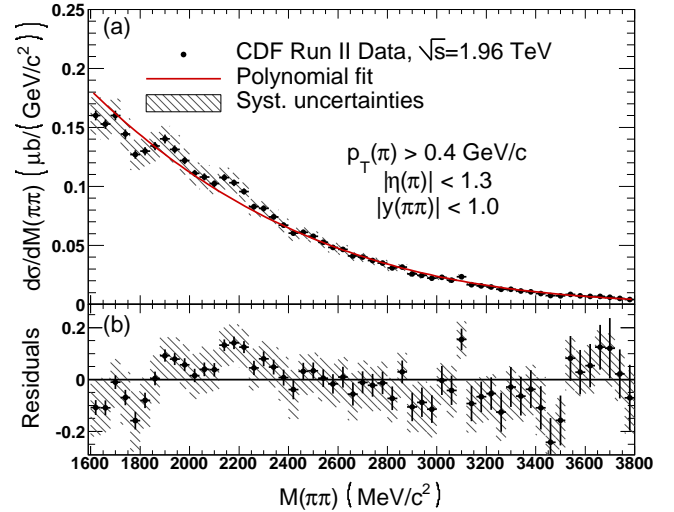}
\caption{(a) Differential cross section at 1.96 TeV in the stated region, with a fit to a fourth-order polynomial. (b)
Relative difference between data and fit as a function of $M(\pi\pi)$. }
\label{fig:Mass1960Tail}
\end{figure}

We first discuss differential cross sections for $M(\pi\pi) >$ 1000 MeV/$c^2$ integrated over $p_T(\pi\pi)$; 
unless otherwise stated we show only $\sqrt{s}$ = 1.96 TeV data, which are more abundant. Figures \ref{fig:Mass1960900} and \ref{fig:MassLog1960} show the differential cross section as a function of $M(\pi\pi)$ and the ratio of the cross sections at $\sqrt{s}$ = 0.9 and 1.96 TeV. The data show a peak centered at 1270 MeV/$c^2$ with a full width at half maximum of approximately 200 MeV/$c^2$, consistent with the $f_2(1270)$ meson. The $f_0(1370)$ may be the cause of the shoulder on the high-mass side of the $f_2(1270)$. An abrupt change
of slope is observed at 1500 MeV/$c^2$, as noted at lower $\sqrt{s}$ \cite{afs,abcdhw} where it is a dip, possibly due to interference between resonances. Structures in the mass distribution are observed up to approximately 2400 MeV/$c^2$, suggesting production of higher-mass resonances. Figure \ref{fig:Mass1960Tail} shows the mass region from 1600 MeV/$c^2$ to 3600 MeV/$c^2$, with a fit to a fourth-order polynomial. There is some structure up to 2400 MeV/$c^2$. Reference \cite{pdg} lists five established resonances above $M$ = 1400 MeV/$c^2$ with seen $\pi\pi$ decays and quantum numbers allowed in \dpe reactions: $f_0(1500),\,f'_0(1525),\,f_0(1710),\,f_2(1950)$, and $f_4(2050)$. The $f_0(1500)$ and the $f_0(1710)$ are both considered to be glueball candidates~\cite{glueball}, but mixing with quarkonium states complicates the issue. From 2400 to 5000 MeV/$c^2$, the data fall monotonically with $M(\pi\pi)$, apart from the small excess at 3100 MeV/$c^2$, which
is consistent with the photoproduction reaction $\gamma + $\pom$ \rightarrow J/\psi \rightarrow e^+e^-$ \cite{cdfchic}. 
 
The differential cross sections at the two energies are similar in shape. The ratio $R(0.9:1.96)$ of the differential cross sections at 0.9 and 1.96 TeV is shown in Fig.~\ref{fig:Mass1960900}(b), and for 1000 $< M(\pi\pi)< $ 2000 MeV it is $R(0.9:1.96) = 1.284\pm0.039$, consistent with
the ratio of approximately 1.3 expected by Regge phenomenology (when both protons are intact), which falls as 1/ln(s) \cite{desai, acf}. However, our data include dissociation, with higher masses $M(p^*)$ allowed at $\sqrt{s} =$ 1.96 TeV, since we require gaps to $\eta = \pm$ 5.9 at both energies. For 2000 $< M(\pi\pi)< $ 3000 MeV $R(0.9:1.96) = 1.560 \pm 0.056$.

\begin{figure}
\centering
\includegraphics[width=86mm]{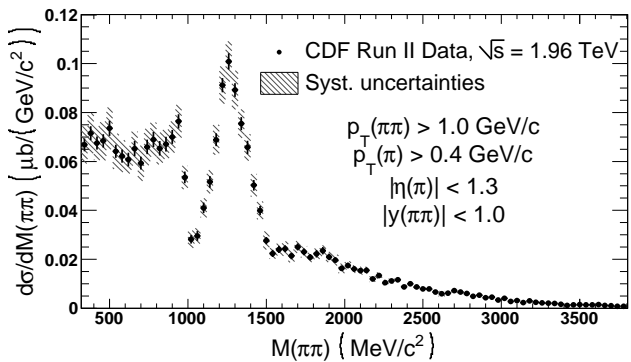}
\caption{As Fig. \ref{fig:Mass1960900} at $\sqrt{s}$ = 1.96 TeV, but with $p_T(\pi\pi) >$ 1.0 GeV/$c$ for which the acceptance
extends to low $M(\pi\pi)$.}
\label{fig:Masshighpt}
\end{figure}

For $p_T(\pi\pi) >$ 1 GeV/$c$ the acceptance extends down to $M(\pi\pi$) = 300 MeV/$c^2$; we show the acceptance-corrected differential cross section in  Fig.~\ref{fig:Masshighpt}. This is approximately uniform up to a sharp drop at $M(\pi\pi)$ = 1000 MeV/$c^2$ seen in previous experiments \cite{acf}, where the $f_0(980)$ and the $K^+K^-$ threshold occur. The absence of a $\rho^0$ signal is expected, as it is forbidden in D$\mathrm{I\!P}$E\nobreakseq{,} and the cross section for photoproduced $\rho^0$-mesons is small, especially for $p_T >$ 1 GeV/$c$. Above $M(\pi\pi$) = 1000 MeV/$c^2$ the same features are observed as in the full sample.  


We previously observed exclusive $\chi_c^0$ production in the $J/\psi+\gamma$ decay channel~\cite{cdfchic}, but the mass resolution was not sufficient to resolve the three $\chi_c$ states, and the $\chi_{c1}$ and $\chi_{c2}$ mesons have much higher branching fractions to the $J/\psi + \gamma$ final state than the $\chi_{c0}(3415)$. If all the $J/\psi + \gamma$ events were attributed to the $\chi_{c0}(3415)$, $d\sigma/dy|_{y=0} = 76 \pm$10(stat)$\pm$10(syst) nb. The $\chi_{c0}(3415)$ decays to $\pi^+\pi^-$ (0.56\%) and $K^+K^-$ (0.61\%), and the $\chi_{c1}$(3510) and $\chi_{c2}$(3556) mesons have smaller branching fractions to these channels. In addition, the CDF~II mass resolution is approximately 25 MeV/$c^2$, less than the mass difference between these states. We do not see significant excesses of events at $M(\pi\pi)$ = 3415 MeV/$c^2$ or at $M(\pi\pi)$ approximately 3280 MeV/$c^2$ where the $K^+K^-$ final state would appear in this distribution.
Using the known  branching fractions \cite{pdg}, efficiency, and $L_{\mathrm{eff}}$, we find $d\sigma/dy|_{y=0} (\chi_{c0}) <$ 35.5 (23.4) nb at 90\% C.L. in the $\pi^+\pi^-$ ($K^+K^-$) decay channels, repectively. These limits imply that $\lesssim$ 50\% of our previous $J/\psi+\gamma$ events were from the $\chi_{c0}(3415)$~\cite{cdfchic}.

Cross section values restricted to the kinematic range of this measurement with $M(\pi\pi) > 2600$ MeV/$c^2$ have been calculated in the DIME Monte Carlo \cite{vkpriv}. There are large uncertainties arising from the unknown $\pi\pi$\pom form factor in this regime, but while this MC with an exponential form factor agrees with the data at 3000 MeV/$c^2$, it predicts a steeper $M(\pi\pi)$ dependence and is lower than the data by a factor of 3 at 5000 MeV/$c^2$. The Orear-like form factor is strongly disfavored, as illustrated in Fig.~\ref{fig:MassLog1960}. We are not aware of any predictions of the cross sections for exclusive $f_0(980), f_2(1270)$ mesons, other low-mass resonances, or cross sections below 2500 MeV/$c^2$.

In summary, we have measured exclusive $\pi^+\pi^-$ production with $|y(\pi\pi)| < 1.0$ and rapidity gaps over $1.3 < |\eta| < 5.9$ in $p\bar{p}$ collisions at $\sqrt{s}$ = 0.9 and 1.96 TeV. The cross section at $\sqrt{s}$ = 1.96 TeV shows a sharp decrease at 1000 MeV/$c^2$ (for $p_T(\pi\pi) > 1$ GeV/$c$), a strong $f_2(1270)$ resonance, and indications of other features  of uncertain origin at higher mass. The cross section at 0.9 TeV is similar in shape, but higher by a factor 1.2 -- 1.6. As the production is expected to be dominated by double pomeron exchange, selecting isospin $I = 0$ and spin $J = 0$ or 2 states, the data can be used to further our knowledge of the isoscalar mesons. We have placed upper limits on exclusive $\chi_{c0}$ production using the $\pi^+\pi^-$ and $K^+K^-$ decay modes. Measurements of \dpe mass spectra in other channels should advance our understanding of scalar and tensor glueballs.

 
We thank the Fermilab staff and the technical staffs of the
participating institutions for their vital contributions. This work
was supported by the US Department of Energy and National Science
Foundation; the Italian Istituto Nazionale di Fisica Nucleare; the
Ministry of Education, Culture, Sports, Science and Technology of
Japan; the Natural Sciences and Engineering Research Council of
Canada; the National Science Council of the Republic of China; the
Swiss National Science Foundation; the A.P. Sloan Foundation; the
Bundesministerium f\"ur Bildung und Forschung, Germany; the Korean
World Class University Program, the National Research Foundation of
Korea; the Science and Technology Facilities Council and the Royal
Society, United Kingdom; the Russian Foundation for Basic Research;
the Ministerio de Ciencia e Innovaci\'{o}n, and Programa
Consolider-Ingenio 2010, Spain; the Slovak R\&D Agency; the Academy
of Finland; the Australian Research Council; and the EU community
Marie Curie Fellowship Contract No. 302103.




\begin{thebibliography}{99}
\bibitem{kine} A coordinate system is used with the $z$-axis along
the proton beam direction; $\theta$ is the polar angle, $\phi$ is the azimuthal
angle, pseudorapidity is $\eta= -\ln \tan(\theta/2)$,
transverse momentum is 
$p_T =|p| \sin\theta$, and transverse energy is $E_T = E \sin\theta$.
\end{thebibliography}
\end{document}